\documentclass[a4paper,10pt,eqno]{article}
\usepackage{theorem}
\usepackage{latexsym,amssymb,amsfonts,amsmath}
\theoremheaderfont{\scshape}

\begin{document}
\title{{\Large {\bf THE FOURIER AND GROVER WALKS ON THE TWO-DIMENSIONAL LATTICE AND TORUS}}}
\author{
{\small Masahiro Asano\footnote{asano-masahiro-kc@ynu.jp},\quad Takashi Komatsu\footnote{komatsu-takashi-fn@ynu.ac.jp},\quad Norio Konno\footnote{konno-norio-bt@ynu.ac.jp},\quad Akihiro Narimatsu\footnote{narimatsu-akihiro-pd@ynu.jp(e-mail of the corresponding author)}}\\
{\scriptsize  Department of Applied Mathematics, Faculty of Engineering, Yokohama National University}\\
{\scriptsize \footnotesize\it 79-5 Tokiwadai, Hodogaya, Yokohama, 240-8501, Japan}\\
}
\date{}
\maketitle
\par\noindent
\begin{small}
\par\noindent
{\bf Abstract}. In this paper, we consider discrete-time quantum walks with moving shift (MS) and flip-flop shift (FF) on two-dimensional lattice $\mathbb{Z}^2$ and torus $\pi_N^2=(\mathbb{Z}/N)^2$. Weak limit theorems for the Grover walks on $\mathbb{Z}^2$ with MS and FF were given by Watabe et al. and Higuchi et al., respectively. The existence of localization of the Grover walks on $\mathbb{Z}^2$ with MS and FF was shown by Inui et al. and Higuchi et al., respectively. Non-existence of localization of the Fourier walk with MS on $\mathbb{Z}^2$ was proved by Komatsu and Tate. Here our simple argument gave non-existence of localization of the Fourier walk with both MS and FF. Moreover we calculate eigenvalues and the corresponding eigenvectors of the $(k_1,k_2)$-space of the Fourier walks on $\pi_N^2$ with MS and FF for some special initial conditions. The probability distributions are also obtained. Finally, we compute amplitudes of the Grover and Fourier walks on $\pi_2^2$. 
\end{small}
\section{Introduction}

The notion of quantum walks (QWs) was introduced by Aharonov et al. \cite{Aharonov1} as a quantum version of the usual random walks. QWs have been intensively studied from various fields. For example, quantum algorithm \cite{Portugal}, the topological insulator \cite{Kitagawa}, and radioactive waste reduction \cite{Matsuoka}.


In the present paper, we consider QWs on the two-dimensional lattice, $\mathbb{Z}^2$, and $N\times N$ torus, $\pi_N^2$ with moving shift (MS) and flip-flop shift (FF). 

The properties of QWs in the one dimension, e.g., ballistic spreading and localization, are well studied, see Konno \cite{Konno}.  On the other hand, the corresponding properties of QWs in the two-dimensions have not been clarified. However, here are some results for the Grover walk and Fourier walk on the two-dimensional lattice. Weak limit theorems for the Grover walks on $\mathbb{Z}^2$ with MS and FF were obtained by Watabe et al. \cite{Watabe et al.} and Higuchi et al. \cite{Higuchi et al.}, respectively. Inui et al. \cite{Inui et al.} and Higuchi et al. \cite{Higuchi et al.} showed that localization occurs for the Grover walks on $\mathbb{Z}^2$ with MS and FF, respectively. Komatsu and Tate \cite{Komatsu and Tate.} proved that localization does not occur for the Fourier walk on $\mathbb{Z}^2$ with MS. It is not known for any result on the Fourier walk on $\mathbb{Z}^2$ with FF. Hence we proved that localization does not occur for the Fourier walk on $\mathbb{Z}^2$ with both MS and FF by using a simple contradiction argument which is different from the method based on the Fourier analysis given by \cite{Komatsu and Tate.}. Time-averaged limit measure of the Grover walk with FF on $\pi_N^2$ was obtained analytically by Marquezino et al. \cite{Marquezino.}. The expression of the time averaged limit measure of the Grover walk on $\pi_N^2$ with both MS and FF is not known. 

In this paper, we compute eigenvalues and the corresponding eigenvectors of the $(k_1,k_2)$-space of the Fourier walks on $\pi_N^2$ with both MS and FF for the special initial conditions, for examples, $k_1=k_2$ or $k_1+k_2\equiv 0\ (\bmod N)$. By using these results, we obtain the measures at time $n$ for the walks. Moreover, we calculate amplitudes of the Fourier walks on $\pi_2^2$ (i.e., $N=2$) with MS and FF. We also compute amplitudes of the Grover walks on $\pi_2^2$ with MS and FF, and discuss a difference between the Fourier and Grover walks.

The rest of the paper is organized as follows. Section 2 is devoted to the definition of QWs on $\pi_N^2$. In Section 3, we consider the Fourier walks on $\pi_N^2$ and $\mathbb{Z}^2$ with both MS and FF. Section 4 deals with the Fourier and Grover walks on $\pi_2^2$. In Section 5, we summarize our results.

\section{QWs\ on\ $\pi_N^2$}
This section presents the definition of QWs on $\pi_N^2$. Let $U$ be a $4\times 4$ unitary matrix which is the coin operator of QW. For a coin operator $U$, we introduce $U_j=P_jU\ (j=1,2,3,4)$, where
\begin{align*}P_1=\begin{bmatrix}1&0&0&0\\0&0&0&0\\0&0&0&0\\0&0&0&0 \end{bmatrix},\ \ P_2=\begin{bmatrix}0&0&0&0\\0&1&0&0\\0&0&0&0\\0&0&0&0 \end{bmatrix},\ \ P_3=\begin{bmatrix}0&0&0&0\\0&0&0&0\\0&0&1&0\\0&0&0&0 \end{bmatrix},\ \ P_4=\begin{bmatrix}0&0&0&0\\0&0&0&0\\0&0&0&0\\0&0&0&1 \end{bmatrix}.\end{align*}
In this paper, we consider two types of shift operator, i.e., moving shift (MS) and flip-flop shift (FF). First we introduce the time evolution of QW with MS as follows: for $x_1,\ x_2\in\pi_N^2$,
\begin{align*}
\Psi_{n+1}^{(m)}(x_1,x_2)&=U_1^{(m)}\Psi_n^{(m)}(x_1+1,x_2)+U_2^{(m)}\Psi_n^{(m)}(x_1-1,x_2)\notag\\&+U_3^{(m)}\Psi_n^{(m)}(x_1,x_2+1)+U_4^{(m)}\Psi_n^{(m)}(x_1,x_2-1),
\end{align*}
where $U^{(m)}=U$. Next we introduce the time evolution of QW with FF as follows:
\begin{align*}
\Psi_{n+1}^{(f)}(x_1,x_2)&=U_1^{(f)}\Psi_n^{(f)}(x_1+1,x_2)+U_2^{(f)}\Psi_n^{(f)}(x_1-1,x_2)\notag\\&+U_3^{(f)}\Psi_n^{(f)}(x_1,x_2+1)+U_4^{(f)}\Psi_n^{(f)}(x_1,x_2-1).
\end{align*}
Here $U^{(f)}$ is given by
\begin{align*}U^{(f)}=\begin{bmatrix} 0&1&0&0 \\ 1&0&0&0\\ 0&0&0&1\\ 0&0&1&0\end{bmatrix}U^{(m)}.\end{align*}
Put $\mathbb{K}_N=\{0,1,\dots, N-1\}$, we define the Fourier transform by
\begin{align*}\Hat{\Psi}_n^{(j)}(k_1,k_2)=\frac{1}{N}\sum_{(x_1,x_2)\in\pi_N^2}\omega^{-(k_1x_1+k_2x_2)}\Psi_n^{(j)}(x_1,x_2)\ \ (j=m,f), \end{align*}
where $(k_1,k_2)\in \mathbb{K}_N^2$\ and $\omega=\exp (2\pi i/N)$.
The time evolution of QW on $(k_1, k_2)$-space is written as
\begin{align*}
\Hat{\Psi}_{n+1}^{(j)}(k_1,k_2)&=U^{(j)}(k_1,k_2)\Hat{\Psi}_{n}^{(j)}(k_1,k_2)\ \ (j=m,f),
\end{align*}
where
\begin{align*}U^{(j)}(k_1,k_2)&=\begin{bmatrix}\omega^{k_1}&0&0&0\\0&\omega^{-k_1}&0&0\\0&0&\omega^{k_2}&0\\0&0&0&\omega^{-k_2}
\end{bmatrix} U^{(j)}\ \ (j=m,f).
\end{align*}
Remark that 
\begin{align}
\Hat{\Psi}_n^{(j)}(k_1,k_2)=\left(U^{(j)}(k_1,k_2)\right)^n\Hat{\Psi}_{0}^{(j)}(k_1,k_2)\ \ (j=m,f).\label{phik}
\end{align}
Since $U^{(j)}(k_1,k_2)$ is unitary, we have the following spectral decomposition as follows:
\begin{align}
U^{(j)}(k_1,k_2)=\sum_{i=0}^{3}\lambda_i^{(j)}(k_1,k_2) |v_i^{(j)}(k_1,k_2)\rangle \langle v_i^{(j)}(k_1,k_2)|\ \ (j=m,f). \label{spec1}
\end{align}
where $\lambda^{(j)}_i(k_1,k_2)$ is eigenvalue of $U^{(j)}(k_1,k_2)$ and $|v_i^{(j)}(k_1,k_2)\rangle$ is the corresponding eigenvector for $i=0,1,2,3$. Therefore, combining Eq.\eqref{phik} with Eq.\eqref{spec1}, we obtain
\begin{align*}
\Hat{\Psi}_n^{(j)}(k_1,k_2)=\sum_{i=0}^{3}\lambda_i^{(j)}(k_1,k_2)^n |v_i^{(j)}(k_1,k_2)\rangle \langle v_i^{(j)}(k_1,k_2)|\Hat{\Psi}^{(j)}_0(k_1,k_2)\ \ (j=m,f). 
\end{align*}
\section{The Fourier\ walk\ on\ $\pi_N^2$}
In this section, we present the definition of the Fourier walks with moving and flip-flop shifts on $\pi_N^2$.
\subsection{The Fourier\ walk with MS}
This subsection deals with the Fourier walk with MS whose coin operator is defined by
\begin{align*}U^{(m)}=\frac{1}{2}\begin{bmatrix}
1&1&1&1\\ 1&i&-1&-i\\1&-1&1&-1\\1&-i&-1&i\end{bmatrix}.\end{align*}
Then $U^{(m)}(k_1,k_2)$ for $k_1,k_2\in\mathbb{K}_N=\{0, 1, \dots, N-1\}$ is given by
\begin{align*}
U^{(m)}(k_1,k_2)=\frac{1}{2}\begin{bmatrix}\omega^{k_1} &\omega^{k_1}&\omega^{k_1}&\omega^{k_1} \\ \omega^{-k_1} &i\omega^{-k_1}&-\omega^{-k_1}&-i\omega^{-k_1} \\ \omega^{k_2} &-\omega^{k_2}&\omega^{k_2}&-\omega^{k_2} \\ \omega^{-k_2} &-i\omega^{-k_2}&-\omega^{-k_2}&i\omega^{-k_2}
\end{bmatrix}.
\end{align*}
Moreover we can compute the characteristic polynomial as follows.
\begin{align}
\det (\lambda I_4-U^{(m)}(k_1,k_2))&=\lambda^4-\frac{1+i}{2}\Bigl(\cos \tilde{k}_1+\sin \tilde{k}_1+\cos \tilde{k}_2+\sin \tilde{k}_2\Bigr)\lambda^3-\frac{1-i}{2}\Bigl(1+\cos \Bigl(\tilde{k}_1-\tilde{k}_2\Bigr)\Bigr)\lambda^2\notag \\
&+\frac{1+i}{2}\Bigl(\cos \tilde{k}_1+\sin \tilde{k}_1+\cos \tilde{k}_2+\sin \tilde{k}_2\Bigr)\lambda-i,\label{chapo}
\end{align}
with $\tilde{k}_j=2\pi k_j/N$. Let $x=\Re(\lambda)$ and $y=\Im(\lambda)$, where $\lambda$ is an eigenvalue of $U^{(m)}(k_1,k_2)$. Here $\Re(z)$ is the real part of $z$ and $\Im(z)$ is the imaginary part of $z\in\mathbb{C}$. We should remark that Eq.\eqref{chapo} implies that $x$ and $y$ satisfy the following equation.
\begin{align}
x^2-y^2-2xy+Ay-B=0, \label{chapoly}
\end{align}
where $A=\cos \tilde{k}_1+\sin \tilde{k}_1+\cos \tilde{k}_2+\sin \tilde{k}_2$\ and\ $B=\{1+\cos (\tilde{k}_1-\tilde{k}_2)\}/2$. It would be difficult to get an explicit form of $\lambda=x+iy$ for any $(k_1,k_2)\in\mathbb{K}_N^2$ by using Eq.\eqref{chapoly}. Therefore we consider the proper subsets ${\cal A}$ of $\mathbb{K}_N^2$. In this model, we deal with the following two cases; $({\bf a})\ {\cal A}=\{(k_1,k_2)=(0,0)\}$ and $({\bf b})\ {\cal A}=\{(k_1,k_2)\in \mathbb{K}_N^2 : k_1=k_2\}$. Let $\lambda_j^{(m)}(k_1,k_2)$ denote the eigenvalues of $U^{(m)}(k_1,k_2)$ and $v^{(m)}_j(k_1,k_2)$ be the corresponding eigenvectors for $j=0,1,2,3$. We should note that case $({\bf a})$ related to a QW starting from uniform initial state given by Eq.\eqref{iniuni}, and case $({\bf b})$ is related to a QW starting from restricted uniform initial state $(x_1+x_2=N)$ given by Eq.\eqref{inikk}.\\ \\
$({\bf a})\ (k_1,k_2)=(0,0)$ case\\
The eigenvalues of $U^{(m)}(0,0)$ are
\begin{align}\lambda_0^{(m)}(0,0)=\lambda^{(m)}_1(0,0)=1,\ \lambda_2^{(m)}(0,0)=-1,\ \lambda_3^{(m)}(0,0)=i,\label{zerozeroei}
\end{align}
and the corresponding eigenvectors are
\begin{align}
v_0^{(m)}(0,0)&=\frac{1}{2}^T\begin{bmatrix}1&1&-1&1\end{bmatrix},\ v_1^{(m)}(0,0)=\frac{1}{\sqrt{2}}^T\begin{bmatrix}1&0&1&0\end{bmatrix}, \notag \\
v_2^{(m)}(0,0)&=\frac{1}{2}^T\begin{bmatrix}1&-1&-1&-1\end{bmatrix},\ v_3^{(m)}(0,0)=\frac{1}{\sqrt{2}}^T\begin{bmatrix}0&1&0&-1\end{bmatrix}. \label{zerozerovec}
\end{align}
$({\bf b})\ (k_1,k_2)=(k,k)$ case\\
The eigenvalues are
\begin{align}
\lambda_0^{(m)}(k,k)=1,\ \lambda_1^{(m)}(k,k)=\omega^k,\ \lambda_2^{(m)}(k,k)=-1,\ \lambda_3^{(m)}(k,k)=i\omega^{-k}, \label{kkei}
\end{align}
and the corresponding eigenvectors are
\begin{align}
v_0^{(m)}(k,k)&=\frac{1}{2}\ {}^T\begin{bmatrix}\omega^k&1&-\omega^k&1 \end{bmatrix},\ v_1^{(m)}(k,k)=\frac{1}{\sqrt{2}}\ {}^T\begin{bmatrix}1&0&1&0 \end{bmatrix}, \notag \\
v_2^{(m)}(k,k)&=\frac{1}{2}^T\begin{bmatrix}\omega^k&-1&-\omega^k&-1 \end{bmatrix},\ v_3^{(m)}(k,k)=\frac{1}{\sqrt{2}}\ {}^T\begin{bmatrix}0&1&0&-1 \end{bmatrix}. \label{kkvec}
\end{align}
From now on, we calculate $\Psi_n^{(m)}(x_1,x_2)$ for two cases. Specific initial states by Eqs.\eqref{zerozeroei},\ \eqref{zerozerovec},\ \eqref{kkei} and \eqref{kkvec}. \\
(i)\ Here we consider with uniform initial state $\Psi_0^{(m)}(x_1,x_2)$, i.e.,
\begin{align}
\Psi_0^{(m)}(x_1,x_2)=\frac{1}{N}\ {}^T\begin{bmatrix} \alpha_1&\alpha_2&\alpha_3&\alpha_4\end{bmatrix}\ ((x_1,x_2)\in \pi_N^2),\label{iniuni}
\end{align}
where $|\alpha_1|^2+|\alpha_2|^2+|\alpha_3|^2+|\alpha_4|^2=1$ with $\alpha_j\in \mathbb{C}\ (j=1,2,3,4)$. By the Fourier transform, we see that the initial state of $(k_1, k_2)$-space becomes 
\begin{align*}\Hat{\Psi}_0^{(m)}(k_1,k_2)=\begin{cases} \cfrac{1}{N}\ {}^T\begin{bmatrix} \alpha_1&\alpha_2&\alpha_3&\alpha_4\end{bmatrix} \ \ (k_1,k_2)=(0,0)\\ \\
{}^T\begin{bmatrix} 0&0&0&0\end{bmatrix}\ \ \ \ \ \ \ \ \ (k_1,k_2)\neq(0,0)
\end{cases}.
\end{align*}
Thus we have 
\begin{align*}
\Psi_n^{(m)}(x_1,x_2)&=\frac{1}{N}\sum_{k_1,k_2 \in \mathbb{K}_N}\omega^{k_1x_1+k_2x_2}\sum_{i=0}^3 \lambda_i(k_1,k_2)^n |v_i(k_1,k_2)\rangle \langle v_i(k_1,k_2)|\phi \rangle \notag \\
&=\frac{1}{N}\sum_{i=0}^3\lambda_i(0,0)^n|v_i(0,0)\rangle \langle v_i(0,0)|\phi \rangle ,
\end{align*}
where $\phi={}^T[\alpha_1\ \alpha_2\ \alpha_3\ \alpha_4]$. From Eqs.\eqref{zerozeroei} and \eqref{zerozerovec}, we obtain the desired result as follows.
\begin{align*}
&\Psi_n^{(m)}(x_1,x_2)\notag \\ &=\frac{1}{4N}\begin{bmatrix}\bigl(3+(-1)^n\bigr)\alpha_1+\bigl(1-(-1)^n\bigr)\alpha_2+(1-(-1)^n\bigr)\alpha_3+\bigl(1-(-1)^n\bigr)\alpha_4 \\
\bigl(1-(-1)^n\bigr)\alpha_1+\bigl(1+(-1)^n+2i^n\bigr)\alpha_2+(-1+(-1)^n)\alpha_3+\bigl(1+(-1)^n-2i^n\bigr)\alpha_4\\
\bigl(1-(-1)^n\bigr)\alpha_1+\bigl(-1+(-1)^n\bigr)\alpha_2+\bigl(3+(-1)^n\bigr)\alpha_3+\bigl(-1+(-1)^n\bigr)\alpha_4\\
\bigl(1-(-1)^n\bigr)\alpha_1+\bigl(1+(-1)^n-2i^n\bigr)\alpha_2+\bigl(-1+(-1)^n\bigr)\alpha_3+\bigl(1+(-1)^n+2i^n\bigr)\alpha_4
\end{bmatrix},
\end{align*}
for $(x_1,x_2)\in \pi_N^2$. Hence we find that the amplitude $\Psi_{n+4}^{(m)}(x_1,x_2)=\Psi_n^{(m)}(x_1,x_2)$ for $(x_1,x_2)\in\pi_N^2$ and $n\in\mathbb{Z}_{\geq}$. That is to say, the Fourier walk with MS starting from uniform initial state has the period $4$.\\
(ii)\ We consider a QW with restricted uniform initial state $\Psi_0^{(m)}(x_1,x_2) $ given by
\begin{align}
\Psi_0^{(m)}(x_1,x_2)=\begin{cases}\cfrac{1}{\sqrt{N}}\ {}^T\begin{bmatrix} \alpha_1&\alpha_2&\alpha_3&\alpha_4\end{bmatrix}\ (x_1+x_2=N)\\ \\ {}^T\begin{bmatrix} 0&0&0&0\end{bmatrix}\ \ \ \ \ \ \ \ \ \ \  (x_1+x_2\neq N)\label{inikk}
\end{cases},
\end{align}
where $|\alpha_1|^2+|\alpha_2|^2+|\alpha_3|^2+|\alpha_4|^2=1$. The Fourier transform implies that the initial state of $(k_1,k_2)$-space becomes
\begin{align*}
\Hat{\Psi}_0^{(m)}(k_1,k_2)=\begin{cases} \cfrac{1}{\sqrt{N}}\ {}^T\begin{bmatrix} \alpha_1&\alpha_2&\alpha_3&\alpha_4\end{bmatrix}\ \ (k_1=k_2)\\ \\ ^T\begin{bmatrix} 0&0&0&0\end{bmatrix}\ \ \ \ \ \ \ \ \ \ \ (k_1\neq k_2)
\end{cases}.
\end{align*}
Therefore we have 
\begin{align}
\Psi_n^{(m)}(x_1,x_2)=\cfrac{1}{N^{3/2}}\sum_{k\in \mathbb{K}_N}\omega^{(x_1+x_2)k}\sum_{i=0}^3 \lambda_i(k,k)^n|v_i(k,k)\rangle \langle v_i(k,k)|\phi \rangle,\label{psikk1}
\end{align}
where $\phi=\ {}^T\begin{bmatrix} \alpha_1&\alpha_2&\alpha_3&\alpha_4\end{bmatrix}$. Combining Eq.\eqref{kkei} with Eq.\eqref{psikk1}, we get 
\begin{align}
\Psi_n^{(m)}(x_1,x_2)=\frac{1}{N^{3/2}}\sum_{k=0}^{N-1} &\omega^{(x_1+x_2)k}\Bigl\{ 1^n\langle v_0^{(m)}(k,k)|\phi\rangle |v_0^{(m)}(k,k)\rangle +\omega^{nk}\langle v_1^{(m)}(k,k)|\phi\rangle |v_1^{(m)}(k,k)\rangle \notag \\
&+(-1)^n\langle v_2^{(m)}(k,k)|\phi\rangle |v_2^{(m)}(k,k)\rangle+i^n\omega^{-nk}\langle v_3^{(m)}(k,k)|\phi\rangle |v_3^{(m)}(k,k)\rangle \Bigr\}.\label{psikk2}
\end{align}
From Eq.\eqref{kkvec}, we see that
\begin{align}
&\langle v_0^{(m)}(k,k)|\phi\rangle=\frac{1}{2}\bigl\{(\alpha_1-\alpha_3)\omega^{-k}+\alpha_2+\alpha_4 \bigr\},\ \ \langle v_1^{(m)}(k,k)|\phi\rangle=\frac{1}{\sqrt{2}}(\alpha_1+\alpha_3),\notag \\
&\langle v_2^{(m)}(k,k)|\phi\rangle=\frac{1}{2}\bigl\{(\alpha_1-\alpha_3)\omega^{-k}-\alpha_2-\alpha_4\bigr\},\ \ \langle v_3^{(m)}(k,k)|\phi\rangle=\frac{1}{\sqrt{2}}(\alpha_2-\alpha_4).\label{vphi}
\end{align}
Inserting Eq.\eqref{vphi} to Eq.\eqref{psikk2} gives
\begin{align}
&\Psi_n^{(m)}(x_1,x_2)=\frac{1}{4N^{3/2}}\sum_{k=0}^{N-1}\omega^{(x_1+x_2)k} \notag\\
&
\begin{bmatrix}
\alpha_1-\alpha_3+(\alpha_2+\alpha_4)\omega^k+2(\alpha_1+\alpha_3)\omega^{nk}+(-1)^n(\alpha_1-\alpha_3)-(-1)^n(\alpha_2+\alpha_4)\omega^k \\
(\alpha_1-\alpha_3)\omega^{-k}+\alpha_2+\alpha_4-(-1)^n\{(\alpha_1-\alpha_3)\omega^{-k}-\alpha_2-\alpha_4 \}+2i^n(\alpha_2-\alpha_4)\omega^{-nk}\\
-\alpha_1+\alpha_3-(\alpha_2+\alpha_4)\omega^k+2(\alpha_1+\alpha_3)\omega^{nk}-(-1)^n\{\alpha_1-\alpha_3-(\alpha_2+\alpha_4)\omega^k\} \\
(\alpha_1-\alpha_3)\omega^{-k}+\alpha_2+\alpha_4-(-1)^n\{(\alpha_1-\alpha_3)\omega^{-k}-\alpha_2-\alpha_4\}-2i^n(\alpha_2-\alpha_4)\omega^{-nk}
\end{bmatrix}. \label{psikk3}
\end{align}
Then Eq.\eqref{psikk3} can be rewritten as
\begin{align*}
&\Psi_n^{(m)}(x_1,x_2)=\frac{1}{4\sqrt{N}}\Biggl\{ \begin{bmatrix}
\alpha_1-\alpha_3\\ \alpha_2+\alpha_4\\ -(\alpha_1-\alpha_3)\notag\\ \alpha_2+\alpha_4 \end{bmatrix}\{1+(-1)^n\}\delta_{j,-j}(x_1,x_2)\notag\\ &+\begin{bmatrix}\alpha_2+\alpha_4\\ 0\\ -(\alpha_2+\alpha_4)\\
 0 \end{bmatrix}\{1+(-1)^n\}\delta_{j,-j-1}(x_1,x_2)+\begin{bmatrix}0\\ \alpha_1-\alpha_3\\ 0\\
 \alpha_1-\alpha_3 \end{bmatrix}\{1-(-1)^n\}\delta_{j,-j+1}(x_1,x_2)\notag\\
&+\begin{bmatrix}\alpha_1+\alpha_3\\0\\ \alpha_1+\alpha_3\\ 0\end{bmatrix}2\delta_{j,-j-n}(x_1, x_2)+\begin{bmatrix}0\\ \alpha_2-\alpha_4\\
0\\-(\alpha_2-\alpha_4)
\end{bmatrix}2i^n\delta_{j,-j+n}(x_1,x_2)\Biggr\} \ \ (j\in\mathbb{K}_N),
\end{align*}
where 
\[\delta_{a,b}(x_1,x_2)=\begin{cases}1\ \bigl((x_1,x_2)=(a,b)\bigr)\\
0\ \bigl((x_1,x_2)\neq (a,b)\bigr)
\end{cases}.
\]
The first, second and third terms of the equation mean that the walker is trapped around $x_1+x_2\equiv0\ (\bmod N)$. The forth and fifth terms of the equation mean that the walker keeps on moving straightly.
\subsection{The Fourier walk with FF}
In this subsection, we consider the Fourier walk with FF whose coin operator is defined by
\begin{align*}U^{(f)}_F=\frac{1}{2}\begin{bmatrix}
1&i&-1&-i\\1&1&1&1\\1&-i&-1&i\\1&-1&1&-1\end{bmatrix}.\end{align*}
Then $U^{(f)}(k_1,k_2)$ for $k_1,k_2\in\mathbb{K}_N=\{0, 1, \dots, N-1\}$ is given by
\begin{align*}
U^{(f)}(k_1,k_2)=\frac{1}{2}\begin{bmatrix}\omega^{k_1} &i\omega^{k_1}&-\omega^{k_1}&-i\omega^{k_1} \\ \omega^{-k_1} &\omega^{-k_1}&\omega^{-k_1}&\omega^{-k_1} \\ \omega^{k_2} &-i\omega^{k_2}&-\omega^{k_2}&i\omega^{k_2} \\ \omega^{-k_2} &-\omega^{-k_2}&\omega^{-k_2}&-\omega^{-k_2}
\end{bmatrix}.
\end{align*}
The eigenvalues of $U^{(f)}(k_1,k_2)$ are the roots of the following polynomial:
\begin{align}
\det (\lambda I_4-U^{(f)}(k_1,k_2))&=\lambda^4-\Bigl(\cos \tilde{k}_1-\cos \tilde{k}_2\Bigr)\lambda^3+\frac{1-i}{2}\Bigl(1-\cos \Bigl(\tilde{k}_1-\tilde{k_2} \Bigr) \Bigr)\lambda^2 \notag\\
&+i\Bigl(\cos \tilde{k}_1-\cos \tilde{k}_2 \Bigr)\lambda-i. \label{chapof}
\end{align}
We put $x=\Re(\lambda)$ and $y=\Im(\lambda)$, where $\lambda$ is an eigenvalue of $U^{(f)}(k_1,k_2)$. We should remark that Eq.\eqref{chapof} implies that $x$ and $y$ satisfy the following equation:
\begin{align*}
x^2-y^2-2xy-C\bigl(x-y\bigr)+D=0,
\end{align*}
where $C=\cos \tilde{k}_1-\cos \tilde{k}_2$\ and\ $D=\{1-\cos (\tilde{k}_1-\tilde{k_2} )\}/2.$ It would be hard to get solution for any $(k_1,k_2)\in\mathbb{K}_N^2$, so we consider suitable proper subsets $\cal{B}$\ $\subset\mathbb{K}_N^2$, as in the case of MS model.\ In this model, we deal with the following two cases: $({\bf a})\ \cal{B}$\ $=\{(k_1,k_2)\in \mathbb{K}_N^2:k_1=k_2\}$ and $({\bf b})\ \cal{B}$\ $=\{(k_1,k_2)\in \mathbb{K}_N^2:k_1+k_2\equiv 0\ \ (\bmod N)\}$. Let $\lambda_j^{(f)}(k_1,k_2)$ be the eigenvalues of $U^{(f)}(k_1,k_2)$ and $v^{(f)}_j(k_1,k_2)$ be the corresponding eigenvectors for $j=0,1,2,3$.\\ \\
$({\bf a})\ (k_1,k_2)=(k,k)$ case\\
The eigenvalues are
\begin{align*}
\lambda_0^{(f)}(k,k)=e^{\pi i/8},\ \lambda_1^{(f)}(k,k)=e^{5\pi i/8}, \ \lambda_2^{(f)}(k,k)=e^{9\pi i/8},\ \lambda_3^{(f)}(k,k)=e^{13\pi i/8},
\end{align*}
and the corresponding eigenvectors are
\begin{align*}
v_j^{(f)}(k,k)=\frac{1}{Z_j(k,k)}\begin{bmatrix}
\omega^k({\lambda_j^{(f)}(k,k)}^2+i\omega^{-2k})(\lambda_j(k,k)+\omega^k)\\
\omega^{-k}({\lambda_j^{(f)}(k,k)}^2+\omega^{2k})(\lambda_j(k,k)+\omega^{-k})\\
-\omega^k({\lambda_j^{(f)}(k,k)}^2+i\omega^{-2k})(\lambda_j(k,k)-\omega^k)\\
\omega^{-k}({\lambda_j^{(f)}(k,k)}^2+\omega^{2k})(\lambda_j(k,k)-\omega^{-k})
\end{bmatrix},
\end{align*}
where $Z_{j}(k,k)$ is a normalized constant.\vspace{0.5\baselineskip}\\
$({\bf b})\ (k_1,k_2)=(k,N-k)$ case\\
The eigenvalues $\lambda$ satisfy the following equation with fourth order;
\begin{align*}
\lambda^4+(1-i)\sin^2\tilde{k}\lambda^2-i=0.
\end{align*}
Thus we get
\begin{align*}
\lambda_j^{(f)}(k,N-k)=&\pm\frac{\sqrt{2-\sin^2\tilde{k}+\sqrt{2-\sin^4\tilde{k}}}+i\sqrt{2+\sin^2\tilde{k}-\sqrt{2-\sin^4\tilde{k}}}}{2},\\ \\
&\pm\frac{\sqrt{2-\sin^2\tilde{k}-\sqrt{2-\sin^4\tilde{k}}}-i\sqrt{2+\sin^2\tilde{k}+\sqrt{2-\sin^4\tilde{k}}}}{2},
\end{align*}
and corresponding eigenvectors are
\begin{align*}
v_j^{(f)}(k,N-k)=\frac{1}{Z_{j}(k,N-k)}
\begin{bmatrix} 
(\lambda_j^{(f)}(k,N-k)+i\sin\Tilde{k})(1+\lambda_j^{(f)}(k,N-k)\omega^k)\\(\overline{\lambda_j^{(f)}(k,N-k)}+i\sin\Tilde{k})(1+\lambda_j^{(f)}(k,N-k)\omega^{-k})\\(\lambda_j^{(f)}(k,N-k)+i\sin\Tilde{k})(1-\lambda_j^{(f)}(k,N-k)\omega^{-k})\\-(\overline{\lambda_j^{(f)}(k,N-k)}+i\sin\Tilde{k})(1-\lambda_j^{(f)}(k,N-k)\omega^{k})
\end{bmatrix},
\end{align*}
where $Z_{j}(k,N-k)$ is a normalized constant.

\subsection{Non-existence of localization}
In this subsection, we prove that localization does not occur for the Fourier walks on $\mathbb{Z}^2$ with both MS and FF. According to Komatsu and Tate \cite{Komatsu and Tate.}, if a QW has localization, then the characteristic polynomial of the quantum coin in $(k_1, k_2)$-space has greater than one constant roots. Assume that the Fourier walk with MS has a constant root $\lambda$ with $|\lambda|=1$. In a similar way, by using characteristic polynomial \eqref{chapo}, we have
\begin{align}
\lambda^4&-\frac{1+i}{2}\Bigl(\cos k_1+\sin k_1+\cos k_2+\sin k_2\Bigr)\lambda^3-\frac{1-i}{2}\Bigl(1+\cos \Bigl(k_1-k_2\Bigr)\Bigr)\lambda^2\notag \\
&+\frac{1+i}{2}\Bigl(\cos k_1+\sin k_1+\cos k_2+\sin k_2\Bigr)\lambda-i=0,\label{chapo0}
\end{align}
where $(k_1,k_2)\in(-\pi,\pi]^2$. Since the constant root $\lambda$ does not depend on $k_1$, we obtain the following equation by differentiating Eq.\eqref{chapo0} with respect to $k_1$.
\begin{align}
-\frac{1+i}{2}\Bigl(-\sin k_1+\cos k_1 \Bigr)\lambda^3+\frac{1-i}{2}\Bigl(\sin\Bigl(k_1-k_2 \Bigr)\Bigr)\lambda^2+\frac{1+i}{2}\Bigl(-\sin k_1+\cos k_1 \Bigr)\lambda=0. \label{bibunms}
\end{align}
Hence Eq.\eqref{bibunms} can be rewritten as
\begin{align}
\Bigl(-\sin k_1+\cos k_1\Bigr)\Bigl(\lambda^2-1\Bigr)-\lambda\sin\Bigl(k_1-k_2 \Bigr)+i\Bigl\{\Bigr(-\sin k_1+\cos k_1\Bigr)\Bigl(\lambda^2-1\Bigr)+\lambda\sin\Bigl(k_1-k_2 \Bigr)\Bigr\}=0.\notag
\end{align}
Therefore we have $\lambda\sin\bigl(k_1-k_2\bigr)=0$ for any $(k_1, k_2)\in(-\pi,\pi]^2$. This contradicts $|\lambda|=1$. Thus we conclude that Eq.\eqref{chapo0} does not have any constant root, so non-existence of localization for the Fourier walk with MS is shown.

In a similar fashion, we can prove that localization does not occur for the Fourier walk with FF. Eq.\eqref{chapof} gives the characteristic polynomial for this model is as follows. 
\begin{align}
\lambda^4-\Bigl(\cos k_1-\cos k_2\Bigr)\lambda^3+\frac{1-i}{2}\Bigl(1-\cos \Bigl(k_1-k_2 \Bigr) \Bigr)\lambda^2+i\Bigl(\cos k_1-\cos k_2 \Bigr)\lambda-i=0,\label{chapof0}
\end{align}
where $(k_1,k_2)\in(-\pi,\pi]^2$. By differentiating Eq.\eqref{chapof0} with respect to $k_1$, we have
\begin{align}
\Bigl(\sin k_1\Bigr)\lambda^3+\frac{1-i}{2}\Bigl(\sin\Bigl(k_1-k_2\Bigr)\Bigr)\lambda^2-i\Bigl(\sin\Tilde{k}_1\Bigr)\lambda=0.\label{bibunf}
\end{align}
Therefore Eq.\eqref{bibunf} becomes
\begin{align}
2\lambda^2\sin k_1+\lambda\sin\Bigl(k_1-k_2\Bigr)-i\Bigl\{2\sin k_1+\lambda\sin\Bigl(k_1-k_2\Bigr) \Bigr\}=0.\notag
\end{align}
Then we obtain $2\sin k_1\pm\sin\Bigl(k_1-k_2\Bigr)=0$ for any $(k_1, k_2)\in(-\pi,\pi]^2$, this is contradiction. Hence we show non-existence of localization for the Fourier walk with FF.

\section{$\pi_2^2$ case}
In this section, we compute $\Psi_n(x_1, x_2)$ of the Fourier and Grover walks with both MS and FF when $N=2$, for all $n\in\mathbb{Z}_{\geq}$ and $(x_1,x_2)\in\pi_2^2$. As the initial state, we take
\begin{align}
\Psi_0^{(j)}(x_1,x_2)=\begin{cases}{}^T\begin{bmatrix} \alpha_1&\alpha_2&\alpha_3&\alpha_4\end{bmatrix}\ \ \bigl((x_1,x_2)=(0,0)\bigr)\\ \\ {}^T\begin{bmatrix} 0&0&0&0\end{bmatrix}\ \ \bigl((x_1,x_2)\neq(0,0)\bigr)\end{cases}\ \ (j=m,f)\notag
\end{align}
for $|\alpha_1|^2+|\alpha_2|^2+|\alpha_3|^2+|\alpha_4|^2=1$ with $\alpha_{\ell}\in \mathbb{C}\ (\ell=1,2,3,4)$.
\subsection{The Fourier walk on $\pi_2^2$}
This subsection deals with $\Psi_n^{(j)}(x_1, x_2)\ \ (j=m,f)$ of the Fourier walk.\\
${\bf(a)}$\ MS case\\
By Eq.\eqref{chapo0}, we get the eigenvalues of $U^{(m)}(k_1,k_2)$ as follows.
\begin{align*}
\lambda_0^{(m)}(0,0)&=1,\ \lambda_1^{(m)}(0,0)=-1,\ \lambda_2^{(m)}(0,0)=1,\ \lambda_3^{(m)}(0,0)=i,\\
\lambda_0^{(m)}(1,1)&=1,\ \lambda_1^{(m)}(1,1)=-1,\ \lambda_2^{(m)}(1,1)=-1,\ \lambda_3^{(m)}(1,1)=-i,\\
\lambda_0^{(m)}(0,1)&=e^{\pi i/8},\ \lambda_1^{(m)}(0,1)=e^{5\pi i/8},\ \lambda_2^{(m)}(0,1)=e^{9\pi i/8},\ \lambda_3^{(m)}(0,1)=e^{13\pi i/8},\\
\lambda_0^{(m)}(1,0)&=e^{\pi i/8},\ \lambda_1^{(m)}(1,0)=e^{5\pi i/8},\ \lambda_2^{(m)}(1,0)=e^{9\pi i/8},\ \lambda_3^{(m)}(1,0)=e^{13\pi i/8},
\end{align*}
and corresponding eigenvectors are
\begin{align*}
v_0^{(m)}(0,0)&=\dfrac{1}{2}\ {}^T
\begin{bmatrix}
1&1&-1&1\\
\end{bmatrix},\ 
v_1^{(m)}(0,0)=\dfrac{1}{2}\ {}^T
\begin{bmatrix}
1&-1&-1&-1\\
\end{bmatrix},\\
v_2^{(m)}(0,0)&=\dfrac{1}{\sqrt{2}}\ {}^T
\begin{bmatrix}
1&0&1&0\\
\end{bmatrix},\ 
v_3^{(m)}(0,0)=\dfrac{1}{\sqrt{2}}\ {}^T
\begin{bmatrix}
0&1&0&-1\\
\end{bmatrix},\\
v_0^{(m)}(1,1)&=-\dfrac{1}{2}\ {}^T
\begin{bmatrix}
1&-1&-1&-1\\
\end{bmatrix},\ 
v_1^{(m)}(1,1)=-\dfrac{1}{2}\ {}^T
\begin{bmatrix}
1&1&-1&1\\
\end{bmatrix},\\
v_2^{(m)}(1,1)&=\dfrac{1}{\sqrt{2}}\ {}^T
\begin{bmatrix}
1&0&1&0\\
\end{bmatrix},\ 
v_3^{(m)}(1,1)=\dfrac{1}{\sqrt{2}}\ {}^T
\begin{bmatrix}
0&1&0&-1\\
\end{bmatrix},\\
v_j^{(m)}(0,1)&=\frac{\sqrt{2(2+\lambda_j^{(m)}+\bar{\lambda}_j^{(m)})}}{4\lambda_j^{(m)}(\lambda_j^{(m)}+1)}
\begin{bmatrix}
\lambda_j(\lambda_j^{(m)}+1)\\
{\lambda_j^{(m)}}^3+1 \\
 \lambda_j^{(m)}(\lambda_j^{(m)}-1)\\
{\lambda_j^{(m)}}^3-1 
\end{bmatrix},\\
v_j^{(m)}(1,0)&=\frac{\sqrt{2(2-\lambda_j^{(m)}-\bar{\lambda}_j^{(m)})}}{4\lambda_j^{(m)}(\lambda_j^{(m)}+1)}
\begin{bmatrix}
\lambda_j^{(m)}(\lambda_j^{(m)}-1)\\
-({\lambda_j^{(m)}}^3+1 )\\
 \lambda_j^{(m)}(\lambda_j^{(m)}+1)\\
-({\lambda_j^{(m)}}^3-1) 
\end{bmatrix}.
\end{align*}
Then we obtain $\Psi_n^{(m)}(x_1,x_2)$ for all $n\in\mathbb{Z}_{\geq}$ and $(x_1,x_2)\in\pi_2^2$.
\begin{align*}
\Psi_{4k}^{(m)}(0,0)=\frac{1+i^k}{2}
\begin{bmatrix}
\alpha_1\\
\alpha_2\\
\alpha_3\\
\alpha_4
\end{bmatrix},\ \Psi_{4k}^{(m)}(1,1)=\frac{1-i^k}{2}
\begin{bmatrix}
\alpha_1\\
\alpha_2\\
\alpha_3\\
\alpha_4
\end{bmatrix},
\end{align*}
\begin{align*}
\Psi_{4k}^{(m)}(0,1)=\Psi_{4k}^{(m)}(1,0)= {}^T
\begin{bmatrix}
0&0&0&0
\end{bmatrix},
\end{align*}
\begin{align*}
\Psi_{4k+1}^{(m)}(0,1)=\dfrac{1}{4}
\begin{bmatrix}
1-i^k & 1-i^k & 1-i^k & 1-i^k \\
1-i^k & i-i^{k+1} & -1+i^k & -i+i^{k+1} \\
1+i^k & -1-i^k & 1+i^k & -1-i^k \\
1+i^k & -i-i^{k+1} & -1-i^k & i+i^{k+1}
\end{bmatrix}
\begin{bmatrix}
\alpha_1\\
\alpha_2\\
\alpha_3\\
\alpha_4
\end{bmatrix},
\end{align*}
\begin{align*}
\Psi_{4k+1}^{(m)}(1,0)=\dfrac{1}{4}
\begin{bmatrix}
1+i^k & 1+i^k & 1+i^k & 1+i^k \\
1+i^k & i+i^{k+1} & -1-i^k & -i-i^{k+1} \\
1-i^k & -1+i^k & 1-i^k & -1+i^k \\
1-i^k & -i+i^{k+1} & -1+i^k & i-i^{k+1}
\end{bmatrix}
\begin{bmatrix}
\alpha_1\\
\alpha_2\\
\alpha_3\\
\alpha_4
\end{bmatrix},
\end{align*}
\begin{align*}
\Psi_{4k+1}^{(m)}(0,0)=\Psi_{4k+1}^{(m)}(1,1)={}^T
\begin{bmatrix}
0&0&0&0
\end{bmatrix},
\end{align*}
\begin{align*}
\Psi_{4k+2}^{(m)}(0,0)=\frac{1}{4}
\begin{bmatrix}
2 & i^k(1+i) & 0 & -i^k(-1+i)\\
i^k(-1+i) & 0 & -i^k(-1+i) & 2\\
0& i^k(-1+i) & 2 & -i^k(1+i)\\
i^k(-1+i)& 2 & -i^k(1+i) & 0
\end{bmatrix}
\begin{bmatrix}
\alpha_1\\
\alpha_2\\
\alpha_3\\
\alpha_4
\end{bmatrix},
\end{align*}
\begin{align*}
\Psi_{4k+2}^{(m)}(1,1)=\frac{1}{4}
\begin{bmatrix}
2 & -i^k(1+i) & 0 & i^k(-1+i)\\
-i^k(-1+i) & 0 & i^k(-1+i) & 2\\
0& -i^k(-1+i) & 2 & i^k(1+i)\\
-i^k(-1+i)& 2 & i^k(1+i) & 0
\end{bmatrix}
\begin{bmatrix}
\alpha_1\\
\alpha_2\\
\alpha_3\\
\alpha_4
\end{bmatrix},
\end{align*}
\begin{align*}
\Psi_{4k+2}^{(m)}(0,1)=\Psi_{4k+2}^{(m)}(1,0)={}^T
\begin{bmatrix}
0&0&0&0
\end{bmatrix},
\end{align*}
\begin{align*}
\Psi_{4k+3}^{(m)}(0,1)=\dfrac{1}{4}
\begin{bmatrix}
1-i^{k+1} & 1-i^{k+1} & 1+i^{k+1} & 1+i^{k+1} \\
1-i^{k+1} & -i-i^k & -1-i^{k+1} & i-i^k \\
1-i^{k+1} & -1+i^{k+1} & 1+i^{k+1} & -1-i^{k+1} \\
1-i^{k+1} & i+i^k & -1-i^{k+1} & -i+i^k
\end{bmatrix}
\begin{bmatrix}
\alpha_1\\
\alpha_2\\
\alpha_3\\
\alpha_4
\end{bmatrix},
\end{align*}
\begin{align*}
\Psi_{4k+3}^{(m)}(1,0)=\dfrac{1}{4}
\begin{bmatrix}
1+i^{k+1} & 1+i^{k+1} & 1-i^{k+1} & 1-i^{k+1} \\
1+i^{k+1} & -i+i^k & -1+i^{k+1} & i+i^k \\
1+i^{k+1} & -1-i^{k+1} & 1-i^{k+1} & -1+i^{k+1} \\
1+i^{k+1} & i-i^k & -1+i^{k+1} & -i-i^k
\end{bmatrix}
\begin{bmatrix}
\alpha_1\\
\alpha_2\\
\alpha_3\\
\alpha_4
\end{bmatrix},
\end{align*}
\begin{align*}
\Psi_{4k+3}^{(m)}(0,0)=\Psi_{4k+3}^{(m)}(1,1)={}^T
\begin{bmatrix}
0&0&0&0
\end{bmatrix},
\end{align*}
where $k\in\mathbb{Z}_{\geq}$.
\vspace{1\baselineskip}\\
${\bf(b)}$\ FF case\\
The eigenvalues are
\begin{align*}
\lambda_0^{(f)}(0,0)&=e^{\pi i/8},\ \lambda_1^{(f)}(0,0)=e^{5\pi i/8},\ \lambda_2^{(f)}(0,0)=e^{9\pi i/8},\ \lambda_3^{(f)}(0,0)=e^{13\pi i/8},\\
\lambda_0^{(f)}(1,1)&=e^{\pi i/8},\ \lambda_1^{(f)}(1,1)=e^{5\pi i/8},\ \lambda_2^{(f)}(1,1)=e^{9\pi i/8},\ \lambda_3^{(f)}(1,1)=e^{13\pi i/8},\\
\lambda_0^{(f)}(0,1)&=1,\ \lambda_1^{(f)}(0,1)=1,\ \lambda_2^{(f)}(0,1)=e^{\pi i/4},\ \lambda_3^{(f)}(0,1)=e^{5\pi i/4},\\
\lambda_0^{(f)}(1,0)&=-1,\ \lambda_1^{(f)}(1,0)=-1,\ \lambda_2^{(f)}(1,0)=-e^{\pi i/4},\ \lambda_3^{(f)}(1,0)=-e^{5\pi i/4},
\end{align*}
and the corresponding eigenvectors are
\begin{align*}
v_j^{(f)}(0,0)&=\dfrac{\sqrt{2(2+\lambda_j^{(f)}+\bar{\lambda}_j^{(f)})}}{4(1+\lambda_j^{(f)})}
\begin{bmatrix}
1+\lambda_j^{(f)} \\
-i{\lambda_j^{(f)}}^2(1+\lambda_j^{(f)})\\
1-\lambda_j^{(f)} \\
i{\lambda_j^{(f)}}^2(1-\lambda_j^{(f)})
\end{bmatrix},\\
v_j^{(f)}(1,1)&=\dfrac{\sqrt{2(2-\lambda_j^{(f)}-\bar{\lambda}_j^{(f)})}}{4(1-\lambda_j^{(f)})}
\begin{bmatrix}
1-\lambda_j^{(f)} \\
-i{\lambda_j^{(f)}}^2(1-\lambda_j^{(f)})\\
1+\lambda_j^{(f)} \\
i{\lambda_j^{(f)}}^2(1+\lambda_j^{(f)})
\end{bmatrix},\\
v_0^{(f)}(0,1)&=\frac{1}{\sqrt{2}}\ {}^T
\begin{bmatrix}
1 & 0 & -1 & 0 \\
\end{bmatrix},\ 
v_1^{(f)}(0,1)=\dfrac{1}{\sqrt{2}}\ {}^T
\begin{bmatrix}
0 & 1 & 0 & 1\\
\end{bmatrix},\\
v_2^{(f)}(0,1)&=\dfrac{1}{2}\ {}^T
\begin{bmatrix}
1  & e^{7\pi i/4} & 1 & -e^{7\pi i/4} \\
\end{bmatrix},\ 
v_3^{(f)}(0,1)=\dfrac{1}{\sqrt{2}}\ {}^T
\begin{bmatrix}
1 & e^{3\pi i/4} & 1 & -e^{3\pi i/4} \\
\end{bmatrix}.
\end{align*}
Thus we have $\Psi_n(x_1,x_2)$ as below.
\begin{align*}
\Psi_{4k}^{(f)}(0,0)=\dfrac{1}{4}
\begin{bmatrix}
2i^k+1+(-1)^k&0&-1+(-1)^k&0\\
0&2i^k+1+(-1)^k&0&1-(-1)^k\\
-1+(-1)^k&0&2i^k+1+(-1)^k&0\\
0&1-(-1)^k&0&2i^k+1+(-1)^k
\end{bmatrix}
\begin{bmatrix}
\alpha_1\\
\alpha_2\\
\alpha_3\\
\alpha_4
\end{bmatrix},
\end{align*}
\begin{align*}
\Psi_{4k}^{(f)}(1,1)=\dfrac{1}{4}
\begin{bmatrix}
2i^k-(1+(-1)^k)&0&-(-1+(-1)^k)&0\\
0&2i^k-(1+(-1)^k)&0&-(1-(-1)^k)\\
-(-1+(-1)^k)&0&2i^k-(1+(-1)^k)&0\\
0&-(1-(-1)^k)&0&2i^k-(1+(-1)^k)
\end{bmatrix}
\begin{bmatrix}
\alpha_1\\
\alpha_2\\
\alpha_3\\
\alpha_4
\end{bmatrix},
\end{align*}
\begin{align*}
\Psi_{4k}^{(f)}(0,1)=\Psi_{4k}^{(f)}(1,0)={}^T
\begin{bmatrix}
0&0&0&0
\end{bmatrix},
\end{align*}
\begin{align*}
\Psi_{4k+1}^{(f)}(0,1)=\dfrac{1}{4}
\begin{bmatrix}
-1+i^k & (-1)^{k+1}i+i^{k+1} & 1-i^k & (-1)^{k}i-i^{k+1} \\
(-1)^{k+1}+i^k & -1+i^k & (-1)^{k+1}+i^k & -1+i^k \\
1+i^k & (-1)^{k+1}i-i^{k+1} & -1-i^k & (-1)^{k}i+i^{k+1} \\
(-1)^{k}+i^k & -1-i^k & (-1)^k+i^k & -1-i^k 
\end{bmatrix}
\begin{bmatrix}
\alpha_1\\
\alpha_2\\
\alpha_3\\
\alpha_4
\end{bmatrix},
\end{align*}
\begin{align*}
\Psi_{4k+1}^{(f)}(1,0)=\dfrac{1}{4}
\begin{bmatrix}
1+i^k & (-1)^{k}i+i^{k+1} & -1-i^k & (-1)^{k+1}i-i^{k+1} \\
(-1)^{k}+i^k & 1+i^k & (-1)^{k}+i^k & 1+i^k \\
-1+i^k & (-1)^{k}i-i^{k+1} & 1-i^k & (-1)^{k+1}i+i^{k+1} \\
(-1)^{k+1}+i^k & 1-i^k & (-1)^{k+1}+i^k & 1-i^k \\
\end{bmatrix}
\begin{bmatrix}
\alpha_1\\
\alpha_2\\
\alpha_3\\
\alpha_4
\end{bmatrix},
\end{align*}
\begin{align*}
\Psi_{4k+1}^{(f)}(0,0)=\Psi_{4k+1}^{(f)}(1,1)={}^T
\begin{bmatrix}
0&0&0&0
\end{bmatrix},
\end{align*}
\begin{align*}
\Psi_{4k+2}^{(f)}(0,0)=\dfrac{1}{4}
\begin{bmatrix}
1+(-1)^ki & 2i^{k+1} & -1+(-1)^ki & 0\\
2i^k & 1+(-1)^ki & 0 & 1-(-1)^ki\\
-1+(-1)^ki & 0&1+(-1)^ki & -2i^{k+1}\\
0 & 1-(-1)^ki & -2i^k & 1+(-1)^ki
\end{bmatrix}
\begin{bmatrix}
\alpha_1\\
\alpha_2\\
\alpha_3\\
\alpha_4
\end{bmatrix},
\end{align*}
\begin{align*}
\Psi_{4k+2}^{(f)}(1,1)=\dfrac{1}{4}
\begin{bmatrix}
-1-(-1)^ki & 2i^{k+1} & 1-(-1)^ki & 0\\
2i^k & -1-(-1)^ki & 0 & -1+(-1)^ki\\
1-(-1)^ki & 0 & -1-(-1)^ki & -2i^{k+1}\\
0 & -1+(-1)^ki & -2i^k & -1-(-1)^ki
\end{bmatrix}
\begin{bmatrix}
\alpha_1\\
\alpha_2\\
\alpha_3\\
\alpha_4
\end{bmatrix},
\end{align*}
\begin{align*}
\Psi_{4k+2}^{(f)}(0,1)=\Psi_{4k+2}^{(f)}(1,0)={}^T
\begin{bmatrix}
0&0&0&0
\end{bmatrix},
\end{align*}
\begin{align*}
\Psi_{4k+3}^{(f)}(0,1)=\dfrac{1}{4}
\begin{bmatrix}
-1+i^{k+1} & (-1)^{k}+i^{k+1} & 1+i^{k+1} & (-1)^{k+1}+i^{k+1} \\
(-1)^{k+1}i+i^k & -1+i^{k+1} & (-1)^{k+1}i-i^k & -1-i^{k+1} \\
1-i^{k+1} & (-1)^{k}+i^{k+1} & -1-i^{k+1} & (-1)^{k+1}+i^{k+1} \\
(-1)^{k}i-i^k & -1+i^{k+1} & (-1)^{k}i+i^k & -1-i^{k+1} \\
\end{bmatrix}
\begin{bmatrix}
\alpha_1\\
\alpha_2\\
\alpha_3\\
\alpha_4
\end{bmatrix},
\end{align*}
\begin{align*}
\Psi_{4k+3}^{(f)}(1,0)=\dfrac{1}{4}
\begin{bmatrix}
1+i^{k+1} & (-1)^{k+1}+i^{k+1} & -1+i^{k+1} & (-1)^{k}+i^{k+1} \\
(-1)^{k}i+i^k & 1+i^{k+1} & (-1)^{k}i-i^k & 1-i^{k+1} \\
-1-i^{k+1} & (-1)^{k+1}+i^{k+1} & 1-i^{k+1} & (-1)^{k}+i^{k+1} \\
(-1)^{k+1}i-i^k & 1+i^{k+1} & (-1)^{k+1}i+i^k & 1-i^{k+1} \\
\end{bmatrix}
\begin{bmatrix}
\alpha_1\\
\alpha_2\\
\alpha_3\\
\alpha_4
\end{bmatrix},
\end{align*}
\begin{align*}
\Psi_{4k+3}^{(f)}(0,0)=\Psi_{4k+3}^{(f)}(1,1)={}^T
\begin{bmatrix}
0&0&0&0
\end{bmatrix}.
\end{align*}
Then we see that the Fourier walks with MS and FF on $\pi_2^2$ have period $16$ i.e., for all $\Psi_{n+16}^{(j)}=\Psi_n^{(j)}\ (j=m,f)$ for $n\in\mathbb{Z}_\geq$. Where $\Psi_n^{(j)}$ is the state of the walk at time $n$.
\subsection{The Grover walk on $\pi_2^2$}
In this subsection, we will check the probability amplitudes of the Grover walk to compare with that of the Fourier walk for the following same initial state: 
\begin{align}
\Psi_0^{(j)}(x_1,x_2)=\begin{cases}{}^T\begin{bmatrix} \alpha_1&\alpha_2&\alpha_3&\alpha_4\end{bmatrix}\ \ \bigl((x_1,x_2)=(0,0)\bigr)\\ \\ {}^T\begin{bmatrix} 0&0&0&0\end{bmatrix}\ \ \bigl((x_1,x_2)\neq(0,0)\bigr)\end{cases}\ \ (j=m,f)\notag
\end{align}
for $|\alpha_1|^2+|\alpha_2|^2+|\alpha_3|^2+|\alpha_4|^2=1$ with $\alpha_{\ell}\in \mathbb{C}\ (\ell=1,2,3,4)$.\\
${\bf(a)}$\ MS case\\
\begin{align*}
\lambda_0^{(m)}(0,0)&=1,\ \lambda_1^{(m)}(0,0)=\lambda_2^{(m)}(0,0)=\lambda_3^{(m)}(0,0)=-1,\\\lambda_0^{(m)}(1,1)&=\lambda_1^{(m)}(1,1)=\lambda_2^{(m)}(1,1)=1,\ \lambda_3^{(m)}(1,1)=-1,\\
\lambda_0^{(m)}(0,1)&=1,\ \lambda_1^{(m)}(0,1)=-1,\ \lambda_2^{(m)}(0,1)=i,\ \lambda_3^{(m)}(0,1)=-i,\\ \lambda_0^{(m)}(1,0)&=1,\ \lambda_1^{(m)}(1,0)=-1,\ \lambda_2^{(m)}(1,0)=i,\ \lambda_3^{(m)}(1,0)=-i,
\end{align*}
the eigenvectors are
\begin{align*}
v_0^{(m)}(0,0)&=\frac{1}{2}\ {}^T
\begin{bmatrix}
1&1&1&1
\end{bmatrix},\ 
v_1^{(m)}(0,0)=\frac{1}{\sqrt{2}}\ {}^T
\begin{bmatrix}
1&-1&0&0
\end{bmatrix},\\
v_2^{(m)}(0,0)&=\frac{1}{\sqrt{2}}\ {}^T
\begin{bmatrix}
0&0&1&-1
\end{bmatrix},\ 
v_3^{(m)}(0,0)=\frac{1}{2}\ {}^T
\begin{bmatrix}
1&1&-1&-1
\end{bmatrix},\\
v_0^{(m)}(1,1)&=\frac{1}{\sqrt{2}}\ {}^T
\begin{bmatrix}
1&-1&0&0
\end{bmatrix},\ 
v_1^{(m)}(1,1)=\frac{1}{\sqrt{2}}\ {}^T
\begin{bmatrix}
0&0&1&-1
\end{bmatrix},\\
v_2^{(m)}(1,1)&=\frac{1}{2}\ {}^T
\begin{bmatrix}
1&1&-1&-1
\end{bmatrix},\ 
v_3^{(m)}(1,1)=\frac{1}{2}\ {}^T
\begin{bmatrix}
1&1&1&1
\end{bmatrix},\\
v_0^{(m)}(0,1)&=\frac{1}{\sqrt{2}}\ {}^T
\begin{bmatrix}
0&0&1&-1
\end{bmatrix},\ 
v_1^{(m)}(0,1)=\frac{1}{\sqrt{2}}\ {}^T
\begin{bmatrix}
1&-1&0&0
\end{bmatrix},\\
v_2^{(m)}(0,1)&=\frac{1}{2}\ {}^T
\begin{bmatrix}
1&1&i&i
\end{bmatrix},\ 
v_3^{(m)}(0,1)=\frac{1}{2}\ {}^T
\begin{bmatrix}
1&1-i&-i
\end{bmatrix},\\
v_0^{(m)}(1,0)&=\frac{1}{\sqrt{2}}\ {}^T
\begin{bmatrix}
1&-1&0&0
\end{bmatrix},\ 
v_1^{(m)}(1,0)=\frac{1}{\sqrt{2}}\ {}^T
\begin{bmatrix}
0&0&1&-1
\end{bmatrix},\\
v_2^{(m)}(1,0)&=\frac{1}{2}\ {}^T
\begin{bmatrix}
1&1&-i&-i
\end{bmatrix},\ 
v_3^{(m)}(1,0)=\frac{1}{2}\ {}^T
\begin{bmatrix}
1&1&i&i
\end{bmatrix}.
\end{align*}

Then we get
\begin{align*}
\Psi_n^{(m)}(0,0)=\dfrac{1+(-1)^n}{8}
\begin{bmatrix}
i^n+3&i^n-1&0&0\\
i^n-1&i^n+3&0&0\\
0&0&i^n+3&i^n-1\\
0&0&i^n-1&i^n+3
\end{bmatrix}
\begin{bmatrix}
\alpha_1\\
\alpha_2\\
\alpha_3\\
\alpha_4\\
\end{bmatrix},
\end{align*}
\begin{align*}
\Psi_n^{(m)}(1,1)=\dfrac{(1-i^n)(1+(-1)^n)}{8}
\begin{bmatrix}
1&1&0&0\\
1&1&0&0\\
0&0&1&1\\
0&0&1&1
\end{bmatrix}
\begin{bmatrix}
\alpha_1\\
\alpha_2\\
\alpha_3\\
\alpha_4\\
\end{bmatrix},
\end{align*}
\begin{align*}
\Psi_n^{(m)}(0,1)=\dfrac{1+(-1)^{n+1}}{8}
\begin{bmatrix}
0&0&1+i^{n+1}&1+i^{n+1}\\
0&0&1+i^{n+1}&1+i^{n+1}\\
1-i^{n+1}&1-i^{n+1}&-2&2\\
1-i^{n+1}&1-i^{n+1}&2&-2
\end{bmatrix}
\begin{bmatrix}
\alpha_1\\
\alpha_2\\
\alpha_3\\
\alpha_4\\
\end{bmatrix},
\end{align*}
\begin{align*}
\Psi_n^{(m)}(1,0)=\dfrac{1+(-1)^{n+1}}{8}
\begin{bmatrix}
-2&2&1-i^{n+1}&1-i^{n+1}\\
2&-2&1-i^{n+1}&1-i^{n+1}\\
1+i^{n+1}&1+i^{n+1}&0&0\\
1+i^{n+1}&1+i^{n+1}&0&0
\end{bmatrix}
\begin{bmatrix}
\alpha_1\\
\alpha_2\\
\alpha_3\\
\alpha_4\\
\end{bmatrix},
\end{align*}

${\bf(b)}$\ FF case\\
\begin{align*}
\lambda_0^{(f)}(0,0)&=\lambda_1^{(f)}(0,0)=\lambda_2^{(f)}(0,0)=1,\ \lambda_3^{(f)}(0,0)=-1,\\
\lambda_0^{(f)}(1,1)&=1,\ \lambda_1^{(f)}(1,1)=\lambda_2^{(f)}(1,1)=\lambda_3^{(f)}(1,1)=-1,\\
\lambda_0^{(f)}(0,1)&=1,\ \lambda_1^{(f)}(0,1)=-1,\ \lambda_2^{(f)}(0,1)=i,\ \lambda_3^{(f)}(0,1)=-i,\\
\lambda_0^{(f)}(1,0)&=1,\ \lambda_1^{(f)}(1,0)=-1,\ \lambda_2^{(f)}(1,0)=i, \lambda_3^{(f)}(1,0)=-i,
\end{align*}
and the corresponding eigenvectors are
\begin{align*}
v_0^{(f)}(0,0)&=\frac{1}{\sqrt{2}}\ {}^T
\begin{bmatrix}
1&-1&0&0
\end{bmatrix},\ 
v_1^{(f)}(0,0)=\frac{1}{\sqrt{2}}\ {}^T
\begin{bmatrix}
0&0&1&-1
\end{bmatrix},\\
v_2^{(f)}(0,0)&=\frac{1}{2}\ {}^T
\begin{bmatrix}
1&1&1&1
\end{bmatrix},\ 
v_3^{(f)}(0,0)=\frac{1}{2}\ {}^T
\begin{bmatrix}
1&1&-1&-1
\end{bmatrix},\\
v_0^{(f)}(1,1)&=\frac{1}{2}\ {}^T
\begin{bmatrix}
1&1&-1&-1
\end{bmatrix},\ 
v_1^{(f)}(1,1)=\frac{1}{\sqrt{2}}\ {}^T
\begin{bmatrix}
1&-1&0&0
\end{bmatrix},\\
v_2^{(f)}(1,1)&=\frac{1}{\sqrt{2}}{}^T
\begin{bmatrix}
0&0&1&-1
\end{bmatrix},\ 
v_3^{(f)}(1,1)=\frac{1}{2}\ {}^T
\begin{bmatrix}
1&1&1&1
\end{bmatrix},\\
v_0^{(f)}(0,1)&=\frac{1}{\sqrt{2}}\ {}^T
\begin{bmatrix}
1&-1&0&0
\end{bmatrix},\ 
v_1^{(f)}(0,1)=\frac{1}{\sqrt{2}}\ {}^T
\begin{bmatrix}
0&0&1&-1
\end{bmatrix},\\
v_2^{(f)}(0,1)&=\frac{1}{2}\ {}^T
\begin{bmatrix}
1&1&i&i
\end{bmatrix},\ 
v_3^{(f)}(0,1)=\frac{1}{2}\ {}^T
\begin{bmatrix}
1&1-i&-i
\end{bmatrix},\\
v_0^{(f)}(1,0)&=\frac{1}{\sqrt{2}}\ {}^T
\begin{bmatrix}
0&0&1&-1
\end{bmatrix},\ 
v_1^{(f)}(1,0)=\frac{1}{\sqrt{2}}\ {}^T
\begin{bmatrix}
1&-1&0&0
\end{bmatrix},\\
v_2^{(f)}(1,0)&=\frac{1}{2}\ {}^T
\begin{bmatrix}
1&1&-i&-i
\end{bmatrix},\ 
v_3^{(f)}(1,0)=\frac{1}{2}\ {}^T
\begin{bmatrix}
1&1&i&i
\end{bmatrix}.
\end{align*}
Thus we have
\begin{align*}
\Psi_n^{(f)}(0,0)=\dfrac{1+(-1)^n}{8}
\begin{bmatrix}
i^n+3&i^n-1&0&0\\
i^n-1&i^n+3&0&0\\
0&0&i^n+3&i^n-1\\
0&0&i^n-1&i^n+3
\end{bmatrix}
\begin{bmatrix}
\alpha_1\\
\alpha_2\\
\alpha_3\\
\alpha_4
\end{bmatrix},
\end{align*}
\begin{align*}
\Psi_n^{(f)}(1,1)=\dfrac{(1-i^n)(1+(-1)^n)}{8}
\begin{bmatrix}
1&1&0&0\\
1&1&0&0\\
0&0&1&1\\
0&0&1&1
\end{bmatrix}
\begin{bmatrix}
\alpha_1\\
\alpha_2\\
\alpha_3\\
\alpha_4
\end{bmatrix},
\end{align*}
\begin{align*}
\Psi_n^{(f)}(0,1)=\dfrac{1+(-1)^{n+1}}{8}
\begin{bmatrix}
0&0&1+i^{n+1}&1+i^{n+1}\\
0&0&1+i^{n+1}&1+i^{n+1}\\
1-i^{n+1}&1-i^{n+1}&2&-2\\
1-i^{n+1}&1-i^{n+1}&-2&2
\end{bmatrix}
\begin{bmatrix}
\alpha_1\\
\alpha_2\\
\alpha_3\\
\alpha_4
\end{bmatrix},
\end{align*}
\begin{align*}
\Psi_n^{(f)}(1,0)=\dfrac{1+(-1)^{n+1}}{8}
\begin{bmatrix}
2&-2&1-i^{n+1}&1-i^{n+1}\\
-2&2&1-i^{n+1}&1-i^{n+1}\\
1+i^{n+1}&1+i^{n+1}&0&0\\
1+i^{n+1}&1+i^{n+1}&0&0
\end{bmatrix}
\begin{bmatrix}
\alpha_1\\
\alpha_2\\
\alpha_3\\
\alpha_4
\end{bmatrix}.
\end{align*}

Compared with the Fourier walks, we see that the Grover walks with both MS and FF on $\pi_2^2$ have period $4$, i.e., $\Psi_{n+4}=\Psi_n$ for $n\in\mathbb{Z}_\geq$. Where $\Psi_n$ is the state of the walk at time $n$.

\section{Summary}
In this paper we considered discrete-time QWs with MS and FF on $\mathbb{Z}^2$ and $\pi_N^2$. We showed that localization does not occur for the Fourier walk on $\mathbb{Z}^2$ with MS and FF by using our contradiction argument which is different from the method based on the Fourier analysis by Komatsu and Tate \cite{Komatsu and Tate.}. Moreover we computed eigenvalues and the corresponding eigenvectors of the $(k_1,k_2)$-space of the Fourier walks on $\pi_N^2$ with MS and FF for some special initial conditions, for instance, $k_1=k_2$ or $k_1+k_2\equiv0\ (\bmod N)$. We derived the measure at time $n$ from these results. In addition, we calculated amplitudes of the Grover and Fourier walks on $\pi_2^2$. One of the interesting future problems would be to obtain the measure at time $n$ of the Fourier walks on $\mathbb{Z}^2$ and $\pi_N^2$ for any initial state.

\end{document}